\pdfoutput=1
\newcommand*{\ATLASLATEXPATH}{latex/}

\documentclass[UKenglish, texlive=2016, cernpreprint]{\ATLASLATEXPATH atlasdoc}

\usepackage{\ATLASLATEXPATH atlaspackage}
\usepackage{\ATLASLATEXPATH atlasbiblatex}
\usepackage{\ATLASLATEXPATH atlascontribute}
\usepackage{\ATLASLATEXPATH atlasphysics}

\usepackage{multirow}

\graphicspath{{logos/}{figures/}}

\usepackage{HIGG-2016-03-PAPER-defs}
\date{\today}

\AtlasTitle{Search for Higgs boson decays into a pair of light bosons 
in the $bb\mu\mu$ final state in $pp$ collision at $\sqrt{s} = $13~\TeV\ with the ATLAS detector}
\author{The ATLAS Collaboration}

\AtlasRefCode{HIGG-2016-03}

\AtlasNote{ATL-COM-PHYS-2018-036}

\PreprintIdNumber{CERN-EP-2018-153}

\AtlasDate{\today}

\AtlasJournal{Phys.\ Lett.\ B.}
\AtlasJournalRef{\PLB 790 (2019) 1}
\AtlasDOI{10.1016/j.physletb.2018.10.073}

\AtlasAbstract{
A search for decays of the Higgs boson into
a pair of new spin-zero particles, $H \rightarrow aa$,  where the $a$-bosons
decay into a $b$-quark pair and a muon pair, is presented. 
The search uses 36.1~\invfb\ of proton--proton collision data at $\sqrt{s}=13$~\TeV\ recorded by the ATLAS
experiment at the LHC in 2015 and 2016.  No significant deviation from the
Standard Model prediction is observed.  Upper limits at  95\%
confidence level are placed on the
branching ratio $(\sigma_H / \sigma_{\textrm{SM}}) \times \mathcal{B}(H\rightarrow aa\rightarrow bb\mu\mu)$, ranging
from $1.2 \times 10^{-4}$ to $8.4\times 10^{-4}$ in the $a$-boson mass range of 20--60~\GeV.
Model-independent limits are set on the visible production cross-section
times the branching ratio to the
$bb\mu\mu$ final state for new physics, $\sigma_{\textrm{vis}}(X) \times
\mathcal{B} (X \rightarrow bb\mu\mu)$, ranging from 0.1~fb to 0.73~fb for 
$m_{\mu\mu}$ between 18 and 62~\GeV.
}

\hypersetup{pdftitle={ATLAS},pdfauthor={The ATLAS Collaboration}}
\DeclareOldFontCommand{\bf}{\normalfont\bfseries}{\mathbf}

\begin{document}
\newif\ifhadhadnote

\maketitle

\section{Introduction}
\label{sec:intro}

The discovery of the Standard Model (SM) Higgs
boson~\cite{HIGG-2012-27,CMS-HIG-12-028} has opened up new avenues
to search for physics beyond the SM (BSM) 
with perspectives to search for non-SM or ``exotic'' decays of the Higgs 
boson. Such searches could provide unique access to hidden-sector
particles that are singlets under the SM gauge transformations~\cite{Curtin:2013fra}. Exotic
decays of the Higgs boson are predicted by many new-physics
models~\cite{Curtin:2013fra,Curtin:2014pda}, including those with an extended Higgs
sector~\cite{Dobrescu:2000yn,Ellwanger:2003jt,Dermisek:2005ar,Chang:2008cw,Morrissey:2008gm}, 
dark matter (DM) models~\cite{Silveira:1985rk,Pospelov:2007mp,Draper:2010ew,Ipek:2014gua,Martin:2014sxa},
models with a first-order electroweak phase
transition~\cite{Profumo:2007wc,Blinov:2015sna} and theories with
neutral naturalness~\cite{Craig:2015pha,Curtin:2015fna}. These
models have also been used to explain the observations of a
$\gamma$-ray excess from the galactic centre (GC) by the Fermi Large
Area Telescope~\cite{TheFermi-LAT:2015kwa,Boehm:2014hva}. For example,
a model for the GC $\gamma$-ray excess was
proposed in which 30~\GeV\ DM particles pair-annihilate dominantly
through a CP-odd scalar
mediator that subsequently decays into SM fermions~\cite{Ipek:2014gua}. If the mediator is
sufficiently lighter than the SM Higgs boson ($H$) then $H$ decay into
the mediator pair can be observed at the LHC.

Existing measurements constrain the 
BSM or ``exotic'' branching ratio ($\mathcal{B}$) of the 125~\GeV\ Higgs boson decays to less
than approximately $34\%$ at $95\%$ confidence level~\cite{HIGG-2015-07}. 
 Due to the narrow width
($\sim4$~\MeV) of the Higgs boson, even a small non-SM coupling of
$\mathcal{O}(10^{-2})$ can lead to $\mathcal{O}(10\%)$ branching ratio
into BSM states. 
This potentially large $\mathcal{B}(H \rightarrow \text{BSM states})$ motivates 
direct searches for exotic $H$ decays.

The analysis presented in this Letter performs the search in
the $bb\mu\mu$ final state.
The $a$-boson can be either a scalar or a pseudoscalar under parity transformations, since the decay mode considered in this search is not sensitive to the difference in coupling.
Assuming that the $a$-boson mixes with the SM Higgs boson and inherits its
Yukawa couplings to fermions, the largest branching ratio is
expected to be to the heaviest fermions accessible by kinematics
(2$m_a<m_H$), where $m_a$ and $m_H$ are the $a$-boson and Higgs boson masses. For
$m_a\gtrsim 10$~\GeV\ this means the $a$-boson would decay preferentially into $bb$.
However, in models with enhanced lepton
couplings such as the Type-III 2HDM~\cite{Branco:2011iw}, the
$a\to\mu\mu$ branching ratio can also be relatively large. Additionally, the
sensitivity of a given channel does not depend only on the expected
signal rate in a particular model, but also on the efficiency for
triggering and reconstructing events of interest. 
The presence of a clean dimuon resonance 
provides a distinctive signature that can be used for triggering and
precision mass reconstruction, which helps to suppress background.

Searches for the Higgs boson with a mass of 125~\GeV\
decaying into two spin-zero particles, $H\rightarrow aa$, 
have been performed in various final states in ATLAS and CMS
~\cite{HIGG-2014-02,CMS-HIG-13-010,CMS-HIG-16-015,EXOT-2016-22,Aaboud:2018gmx,Sirunyan:2018pzn,HIGG-2017-05}.
The CMS search with $\sqrt{s}=8$~\TeV\ data in the $bb\mu\mu$ final state set 95\% CL limits on $(\sigma_H / \sigma_{\textrm{SM}}) \times \mathcal{B}(H\rightarrow aa\rightarrow bb\mu\mu)$ between $2 \times 10^{-4}$ and $8\times 10^{-4}$ in the $a$-boson mass range of 25--62.5~\GeV~\cite{CMS-HIG-16-015}. In Type-III 2HDM+S scenario with tan $\beta$~=~2~\cite{Curtin:2014pda}, where tan~$\beta$ denotes the ratio of the vacuum expectation values of the two Higgs fields,  these limits translate into upper limits on $(\sigma_H / \sigma_{\textrm{SM}}) \times \mathcal{B}(H\rightarrow aa)$ ranging between 13\% and 50\%. Some of the most stringent limits up to date for Type-III 2HDM+S with tan $\beta$~=~2 come from the CMS search with $\sqrt{s}=13$~\TeV\ data in the $bb\tau\tau$ final state, setting the upper limits on $(\sigma_H / \sigma_{\textrm{SM}}) \times \mathcal{B}(H\rightarrow aa)$ between 4\% and 26\% in the $a$-boson mass range of 15--60~\GeV~\cite{Sirunyan:2018pzn}.

\section{Data and simulation}
\label{sec:samples}

The search presented in this Letter is based on the 36.1~$\invfb$ dataset of proton--proton 
collisions at a centre-of-mass energy of $\sqrt{s}$~= 13~\TeV\
recorded by the ATLAS experiment at the LHC during 2015 and 2016.
The ATLAS experiment~\cite{PERF-2007-01}  is a
multipurpose particle detector with a forward--backward symmetric
cylindrical geometry and a near $4\pi$ coverage in solid angle.\footnote{
  The ATLAS Collaboration uses a right-handed coordinate system with
  its origin at the nominal interaction point (IP) in the centre of
  the detector and the $z$-axis along the beam pipe. The $x$-axis
  points from the IP to the centre of the LHC ring, and the $y$-axis
  points upwards. Cylindrical coordinates $(r,\phi)$ are used in the
  transverse plane, $\phi$ being the azimuthal angle around the beam
  pipe. The pseudorapidity is defined in terms of the polar angle
  $\theta$ as $\eta=-\ln\tan(\theta/2)$. Angular distance is  measured
  in units of $\Delta R \equiv  \sqrt{(\Delta\eta)^2 + (\Delta\phi)^2}
  $.} It consists of an inner tracking detector surrounded by a thin
superconducting solenoid providing a 2~T axial magnetic field,
electromagnetic and hadronic calorimeters, and a muon
spectrometer. Events are collected with single-muon triggers requiring the muon \pt\ to be above 24 or 26~\GeV, depending on the
data-taking period.
The trigger efficiency for the signal events with the muon \pt\ on the trigger plateau is about 80\%. 

Simulated events are used to model the signal and SM backgrounds
processes. Higgs boson production through the  gluon--gluon fusion (ggF) and vector-boson fusion (VBF) processes was
modelled at next-to-leading order (NLO) using \textsc{Powheg-Box}
v2~\cite{Nason:2004rx,Frixione:2007vw,Alioli:2010xd} interfaced with \textsc{Pythia}
8.186~\cite{PYTHIA82} using the AZNLO set of tuned parameters~\cite{STDM-2012-23}
for the simulation of the $bb\mu\mu$ decay of the Higgs boson, as well
as for parton showering and hadronisation. The ggF Higgs boson production
rate is normalised to the total cross-section predicted by a
next-to-next-to-next-to-leading-order QCD calculation with NLO electroweak corrections applied~\cite{deFlorian:2016spz,Anastasiou:2015ema,Anastasiou:2016cez,Actis:2008ug,Anastasiou:2008tj}.
The VBF production rate is normalised to an approximate next-to-next-to-leading-order (NNLO) QCD cross-section with NLO electroweak corrections applied~\cite{Heinemeyer:2013tqa,Ciccolini2007,Ciccolini2008,BolzoniVBF}.
Five mass points were simulated in the range $m_a = \text{20--60}~\GeV$ in
steps of $10~\GeV$ for both ggF and VBF production. 

\textsc{Sherpa} 2.2.1~\cite{Gleisberg:2008ta} with the
\textsc{NNPDF3.0}~\cite{Ball:2014uwa} set of parton distribution functions (PDF)
was  used for the generation of Drell--Yan, $W$+jets and diboson
($WW$, $WZ$, $ZZ$) backgrounds.
Cross-sections were calculated at NNLO QCD accuracy for $Z^{(*)}/\gamma^{*}$+jets
and $W$+jets production~\cite{WZcrossSections2} and at NLO including LO
contributions with two additional partons for the diboson processes~\cite{Gleisberg:2008ta,Hoche:2010kg,Campbell:2011bn}.
The $t\bar{t}$ and single-top-quark samples were generated
with \textsc{Powheg-Box} v2~\cite{Frixione:2007vw} using the \textsc{CT10} PDF set~\cite{Lai:2010vv}
interfaced with \textsc{Pythia} v6.428~\cite{Sjostrand:2006za} and the Perugia 2012 
set of tuned parameters~\cite{Skands:2010ak} for the parton shower.
The mass of the top quark ($m_{t}$) was set to 172.5~\GeV. The parameter
$h_{\textrm{damp}} $ in \textsc{Powheg}, used to regulate the 
high-\pt\ radiation, was set to $m_{t}$ for improved agreement between 
data and simulation in the high \pt\ region~\cite{ATL-PHYS-PUB-2014-005}.  The
cross-section of $t\bar{t}$ was calculated at NNLO in QCD
including resummation of next-to-next-to-leading logarithmic (NNLL) soft gluon
terms~\cite{Czakon:2013goa,Czakon:2011xx}. 
The cross-section for single-top-quark production was calculated with the prescriptions in
Refs.~\cite{Kidonakis:2011wy, Kidonakis:2010ux}.  The production of
$\ttbar$ pairs in association with  $W$/$Z$ bosons (denoted by $\ttbar V$) was modelled with
samples generated at LO using \textsc{MadGraph5\_aMC@NLO}
v2.2.2~\cite{Alwall:2014hca} and showered with \textsc{Pythia}
v8.186. The samples are normalised to NLO cross-sections~\cite{Lazopoulos:2008de,Campbell:2012dh}.

Additional $pp$ collisions generated with \textsc{Pythia} v8.186 were overlaid to
model the effects of additional interactions in the same and neighbouring bunch
crossings (pile-up) for all simulated events. The pile-up simulation
used the A2 set of tuned parameters~\cite{ATL-PHYS-PUB-2012-003} and the MSTW2008LO PDF set~\cite{Martin:2009iq}.
All the samples were processed through the full ATLAS detector simulation~\cite{SOFT-2010-01} based on
\textsc{GEANT4}~\cite{Agostinelli:2002hh} and processed with the same reconstruction algorithm as used for data.

\section{Selection criteria}
\label{sec:reconstruction}

Interaction vertices from proton--proton collisions are
reconstructed from at least two tracks with transverse momentum (\pt) larger than 0.4~\GeV, and are
required to be consistent with the beamspot envelope. The primary
vertex (PV) is identified as the one with the largest $\sum p_{\textrm{T}}^2$ of associated tracks~\cite{ATL-PHYS-PUB-2015-026}.

\label{subsec:reco_muons}

Muon candidates are reconstructed using the information from the inner
detector and the muon spectrometer~\cite{PERF-2015-10}.  
They are required to satisfy ``medium'' identification
criteria~\cite{PERF-2015-10}, be matched to the PV and have $\pt>7$~\GeV\ and $|\eta|<2.7$.
Additionally, the muons must satisfy the following criteria: 
the projected longitudinal impact
parameter $|z_0\sin\theta|$ must be less than 0.5~mm and the ratio of the transverse impact
parameter $d_0$ to its estimated uncertainty $\sigma_{d_0}$,
$|d_0/\sigma_{d_0}|$, must be less than 3.
Finally, the selected muons must fulfil requirements on the scalar sum of
\pt\ of additional inner detector tracks and on the 
sum of the \et\ of calorimeter topological clusters~\cite{PERF-2014-07} in a cone of size $\Delta R = 0.2$ around the muon
to ensure they satisfy ``tight'' isolation criteria~\cite{PERF-2015-10}.
These requirements select signal muons with an identification efficiency of $\sim$94\%
 and isolation efficiency ranging between $\sim$91\% for $m_a = 20~\GeV$  and $\sim$95\% for $m_a = 60~\GeV$.

\label{subsec:reco_calojets}

Jets are reconstructed using the anti-$k_t$
algorithm~\cite{Cacciari:2008gp} implemented in the \textsc{FastJet}\xspace package~\cite{Cacciari:2011ma} with a radius parameter $R = 0.4$
applied to topological clusters of energy deposits in calorimeter cells. 
 Jets from pile-up are suppressed with the use of tracking information as detailed in
Ref.~\cite{PERF-2014-03}.  All selected jets are
required to have $\pt > 20$~\GeV, $|\eta| <2.5$ and must pass quality
requirements defined to minimise the impact of detector effects, beam backgrounds and
cosmic rays.

\label{subsec:reco_btag}

Jets consistent with the hadronisation of a $b$-quark ($b$-jets) are identified
using a multivariate
discriminant~\cite{PERF-2012-04,ATL-PHYS-PUB-2016-012}.
 This analysis uses the 77\% $b$-jet identification efficiency working
 point 
for which the purity of the $b$-tagged sample is approximately
95\%, while 
the probability of misidentifying a jet initiated by a charm quark as a $b$-jet is
approximately 16\%, as determined from a sample of simulated \ttbar\ events.   

In order to reject non-prompt muons from the decay of hadrons within a jet,
an overlap removal algorithm is applied. 
If a jet is found within $\Delta R = 0.4$ of the muon candidate, the
overlap is resolved in the following way: 
if there are more than two tracks with $\pt>500$~\MeV\  associated with the jet then the muon is removed from the event, otherwise the muon is retained and the jet is removed.

\label{subsec:reco_met}
The missing transverse momentum (\met) used in the analysis is
calculated as the magnitude of the negative vector sum ($\overrightarrow{p}_{\textrm{T}}^{\textrm{miss}}$) of the transverse momenta of all selected and calibrated
objects in the event  and the additional
``soft'' term that takes into account 
tracks not associated with any of the these objects~\cite{Aaboud:2018tkc}. 
The ``soft'' term is calculated from
inner detector tracks matched to the PV and included to achieve a
better \met\ resolution. 

Events are required to have exactly two $b$-tagged jets with $\pt>20$~\GeV\ and exactly two reconstructed muons of opposite charge, with
the leading muon having $\pt > 27~\GeV$ to 
be in the maximum-efficiency regime of the trigger and the subleading muon having $\pT>7~\GeV$. 
The dimuon invariant mass ($m_{\mu\mu}$) is required to be between 16~\GeV\ and 64~\GeV. The upper bound on $m_{\mu\mu}$ is defined by the assumption that the 125~\GeV\ Higgs boson decays into two on-shell particles of equal masses, while the lower bound is motivated by the kinematics of the $a$-boson decays. For lower values of $m_a$, most of the signal jets fall below the reconstruction threshold and the jets tend to overlap geometrically in the detector so that the sensitivity of the analysis to the $H\rightarrow aa$ signal decreases.
This set of selection criteria is referred to as the ``preselection''.

\begin{figure}
  \centering
  \subfloat[a][]{
    \includegraphics[width=0.5\textwidth]{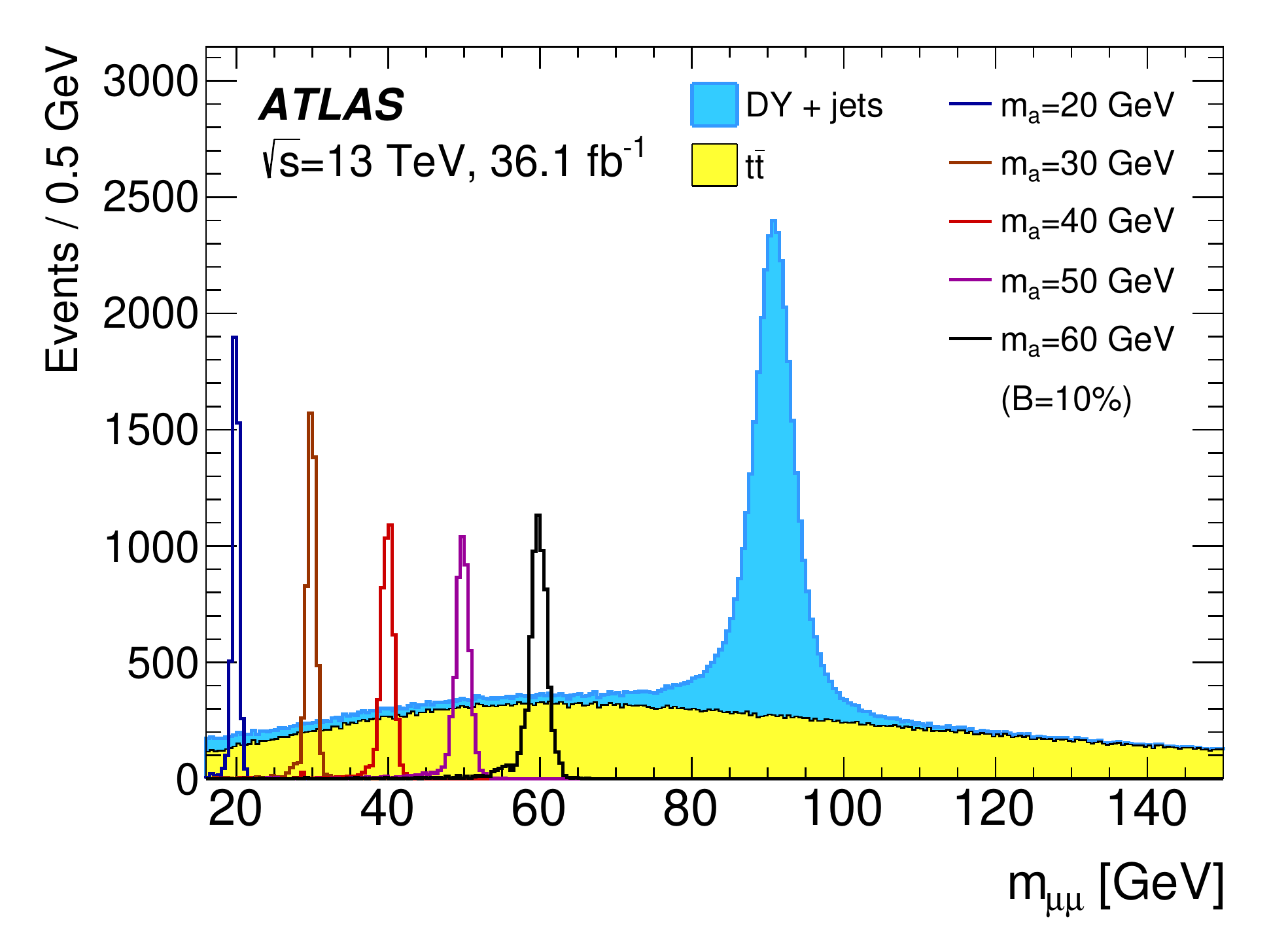}
  }
  \subfloat[b][]{
    \includegraphics[width=0.5\textwidth]{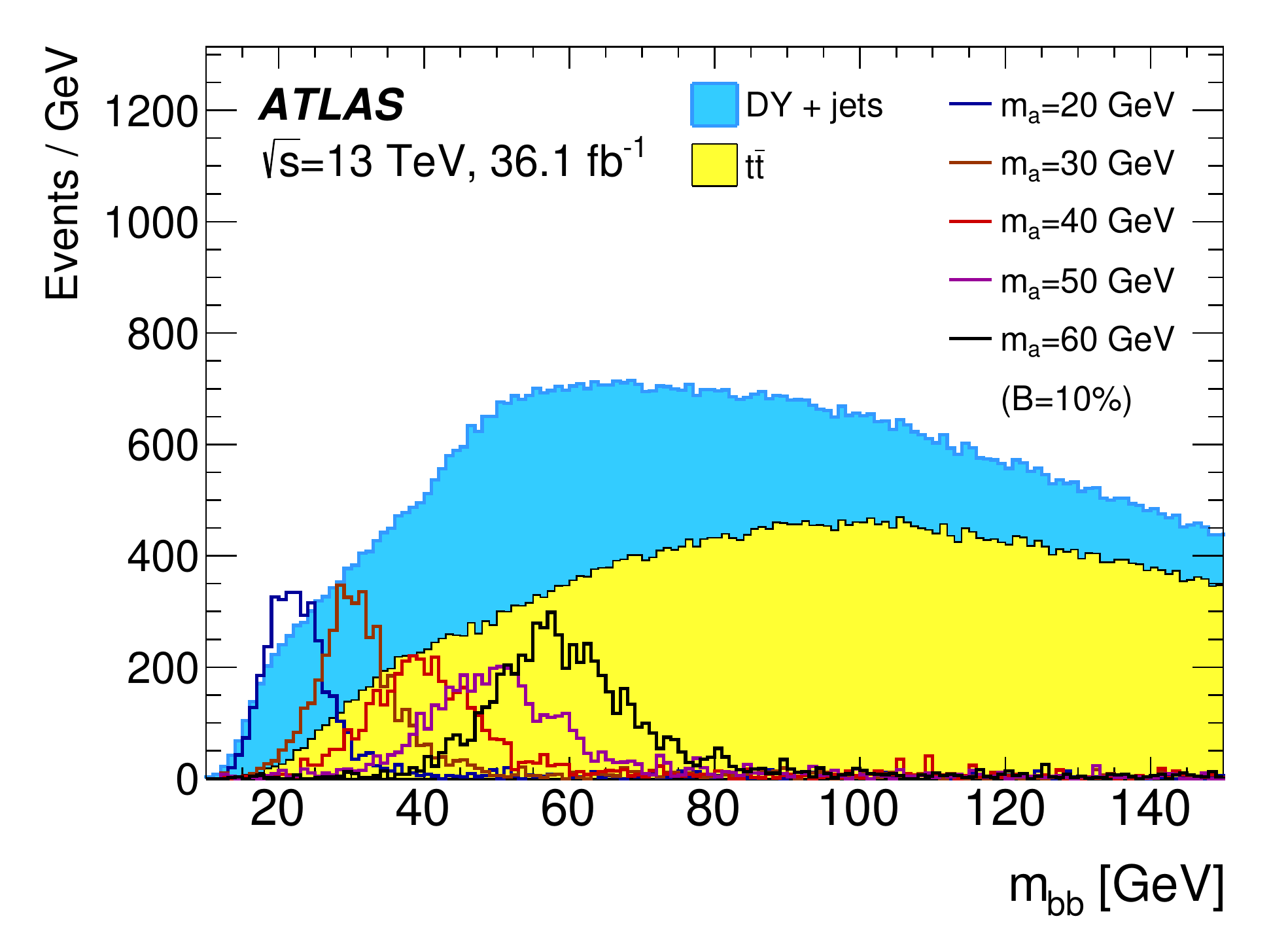}
  }

  \subfloat[c][]{
    \includegraphics[width=0.5\textwidth]{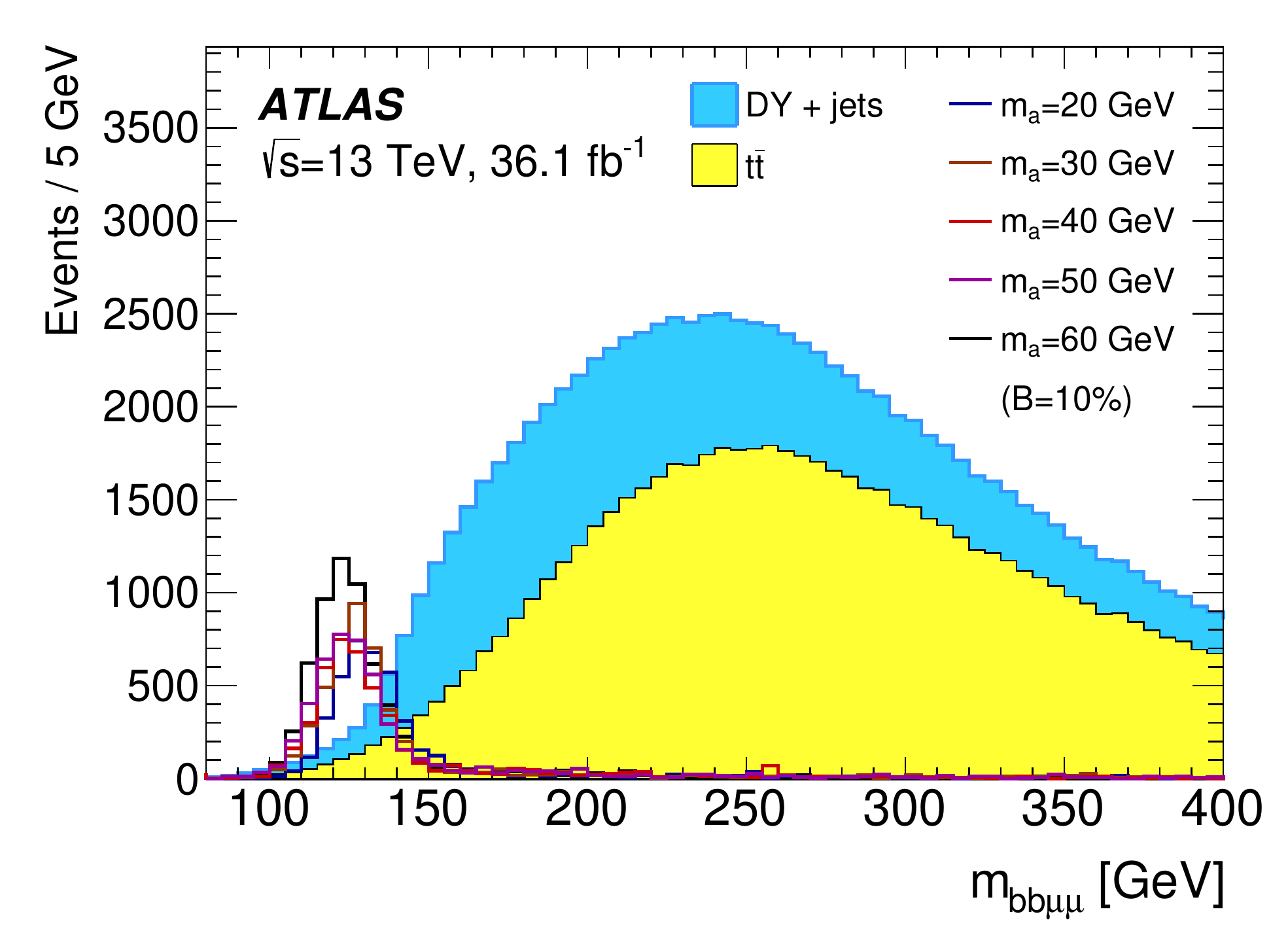}
  }
  \subfloat[d][]{
    \includegraphics[width=0.5\textwidth]{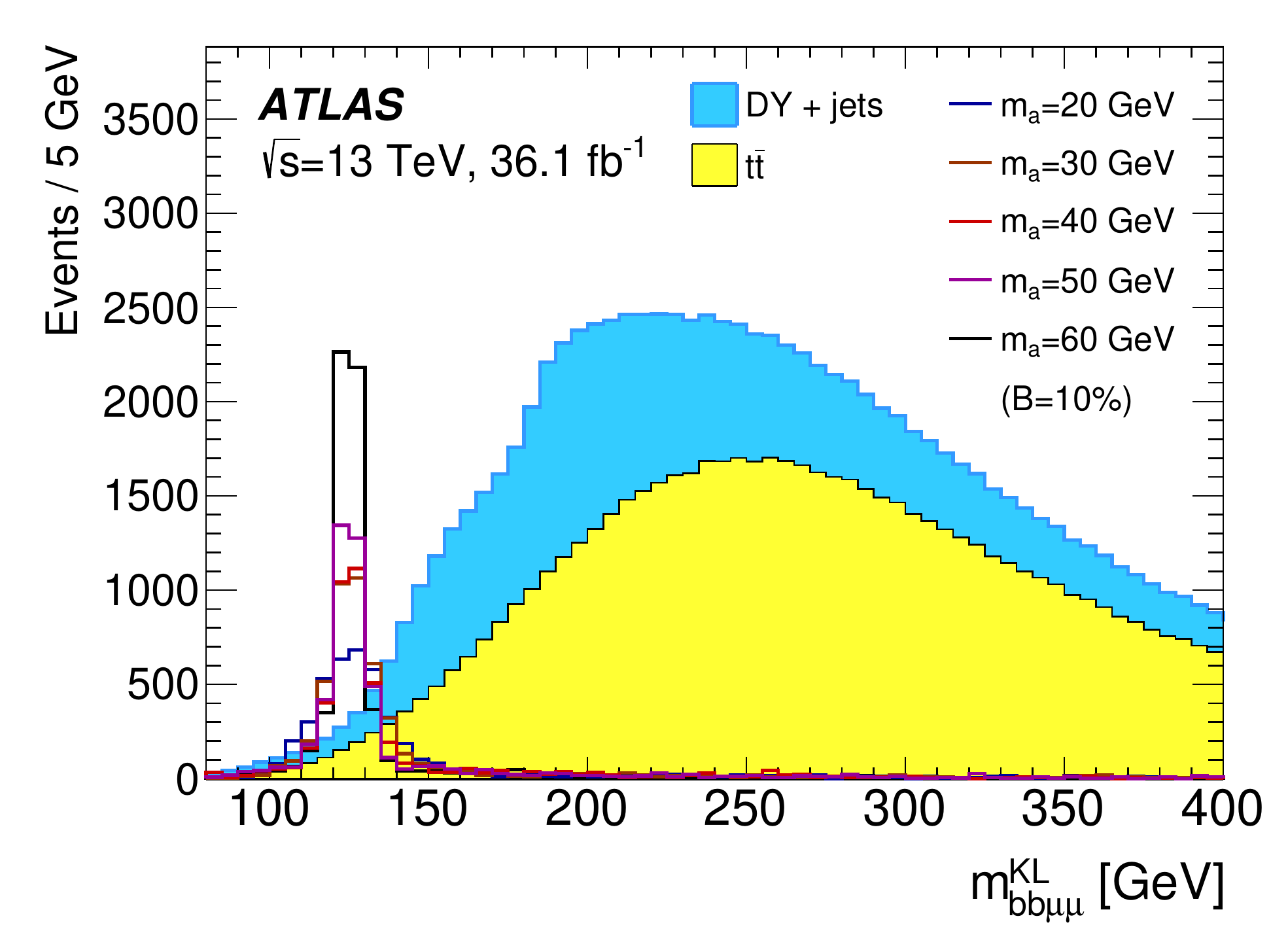}
  }
  \caption{
  The (a) $m_{\mu\mu}$, (b) $m_{bb}$ before the KL fit, (c) $m_{bb\mu\mu}$ before and (d) $m_{bb\mu\mu}^{\textrm{KL}}$ after the KL fit for events after the preselection stage, but removing the upper bound on $m_{\mu\mu}$.  The \ttbar\ contribution is modelled with the simulated sample normalised to the theoretical cross-section. The Drell--Yan contribution is taken from data templates (described in the text) and normalised to the total yield predicted by the Drell--Yan simulation. The signal distributions for all five simulated $m_a$ are also shown assuming the SM Higgs boson cross-section (including ggF, VBF and $VH$ production) and $\mathcal{B}(H \rightarrow aa \rightarrow bb\mu\mu)=10\%$. The branching ratio in this and all subsequent figures is chosen so as to give good visibility on the plot.
  }
  \label{fig:mbb_muu}
\end{figure}

Signal events are characterised by the invariant mass of the two $b$-jets ($m_{bb}$) being equal, within the detector resolution, to the dimuon invariant mass and the four-object mass ($m_{bb\mu\mu}$) being approximately 125~\GeV. One side of the $H\rightarrow aa$ decay ($a\rightarrow\mu\mu$) is measured with
approximately ten times better resolution than the other side of the decay
($a\rightarrow bb$), as shown in Figures~\ref{fig:mbb_muu}(a) and~\ref{fig:mbb_muu}(b). 

A kinematic-likelihood (KL) fit~\cite{Erdmann:2013rxa} exploiting the symmetry of $H\rightarrow aa$ decays is performed to test the compatibility of an event with the $m_{bb} \simeq m_{\mu\mu}$ hypothesis and improve the $m_{bb\mu\mu}$ resolution in signal events.
The KL fit finds the energies of the leading ($\hat{E}_{b_1}$) and subleading
($\hat{E}_{b_2}$) $b$-jets that maximise the likelihood for an event with
measured leading and subleading $b$-jet energies $E_{b_1}$ and
$E_{b_2}$ and with dimuon invariant mass $m_{\mu\mu}$. The
likelihood is defined as follows, 
\begin{equation*}
L = W(\hat{E}_{b_1},E_{b_1})\cdot W(\hat{E}_{b_2},E_{b_2})\cdot F_\textrm{BW}(m_{bb}^{\textrm{KL}},m_{\mu\mu}), 
\end{equation*}
where $m_{bb}^{\textrm{KL}}$ is the dijet invariant mass computed from the $b$-jet
four-momenta corresponding to $\hat{E}_{b_1}$ and $\hat{E}_{b_2}$, $W$
is the transfer function of the $b$-jets, and $F_\textrm{BW}$ is a
Breit--Wigner function centred on $m_{\mu\mu}$ with a width that is small compared to the $m_{bb}$ resolution.
The transfer function
$W(\hat{E}_{b_1},E_{b_1})$ is a double Gaussian probability density
function derived from simulated events as a function of jet \pt and $\eta$ using the difference between true and reconstructed energies. 
The fit determines a maximum-likelihood value of $L$ (denoted by $\textrm{ln}(L^\textrm{max})$), which quantifies how well 
an event fits to the constraints. The $b$-jet momenta determined by the fit are used to recompute the
four-body mass denoted $m_{bb\mu\mu}^{\textrm{KL}}$.
As seen in Figure~\ref{fig:mbb_muu}(d), the resolution of the $m_{bb\mu\mu}^{\textrm{KL}}$ distribution for the signal is improved by up to a factor of two compared to the pre-fit $m_{bb\mu\mu}$ shown in Figure~\ref{fig:mbb_muu}(c), while the background shape within the $m_{bb\mu\mu}$ signal peak remains almost unchanged with the yields rising by $\sim$20\%. This allows the analysis to place tighter constraints on the difference between the reconstructed invariant mass of the $bb\mu\mu$ system and $m_H$, rejecting more background events and obtaining higher signal significance.

Two criteria based on the kinematic likelihood fit are applied to select signal-like events and reject background events that do not fit the $m_{bb} = m_{\mu\mu}$ constraint well: $|m_{bb\mu\mu}^{\textrm{KL}} - m_{H}| < 15$~\GeV\ and $\ln(L^\textrm{max}) > -8$. Finally, the $\MET < 60$~\GeV\ requirement rejects a large portion of \ttbar~pairs where both top quarks decay semileptonically, while retaining most of the signal events.
Adding these three requirements after the preselection stage defines the signal-enhanced region (SR). 
A search for a localised excess above the expected background 
is performed in multiple $m_{\mu\mu}$ bins of the SR centred around the hypothesised $m_a$.
A bin width of 2~\GeV\ is chosen for $16 < m_{\mu\mu} \leq 40$~\GeV, 3~\GeV\ for  $40< m_{\mu\mu} < 50$~\GeV\ and 
4~\GeV\ for  $50\leq m_{\mu\mu} < 64$~\GeV\, respectively, in order to maximise the sensitivity.

\section{Backgrounds}
\label{sec:background}

The dominant backgrounds in the signal region are Drell--Yan (DY) dimuon events in association with $b$-quarks and  pair production of top quarks where both $W$ bosons from top quarks decay into muons. Each of the dominant backgrounds amounts to approximately 50\% of the total background in the SR. 
Two control regions (CR) are defined to constrain the contributions of the dominant backgrounds in the signal region.
 They are chosen such that they have negligible signal contamination, but are kinematically close to the SR to reduce model dependence.  The top control region (TCR) is defined by applying the same selection criteria as for the signal region, but inverting the requirement on the missing transverse momentum to $\MET>60$~\GeV.  
According to the simulation, approximately 95\% of the events in TCR originate from \ttbar~production. The Higgs boson mass sidebands of the signal region are used as the Drell--Yan control region (DYCR): the constraint on the $bb\mu\mu$ invariant mass after the KL fit  is inverted to $80 < m_{bb\mu\mu}^{\textrm{KL}} < 110$~\GeV\ or $140 < m_{bb\mu\mu}^{\textrm{KL}} < 170$~\GeV. The DYCR consists of about 50\% DY events and about 50\% \ttbar\ events.

\begin{figure}[!htbp]
\centering
  \includegraphics[width=0.49\textwidth]{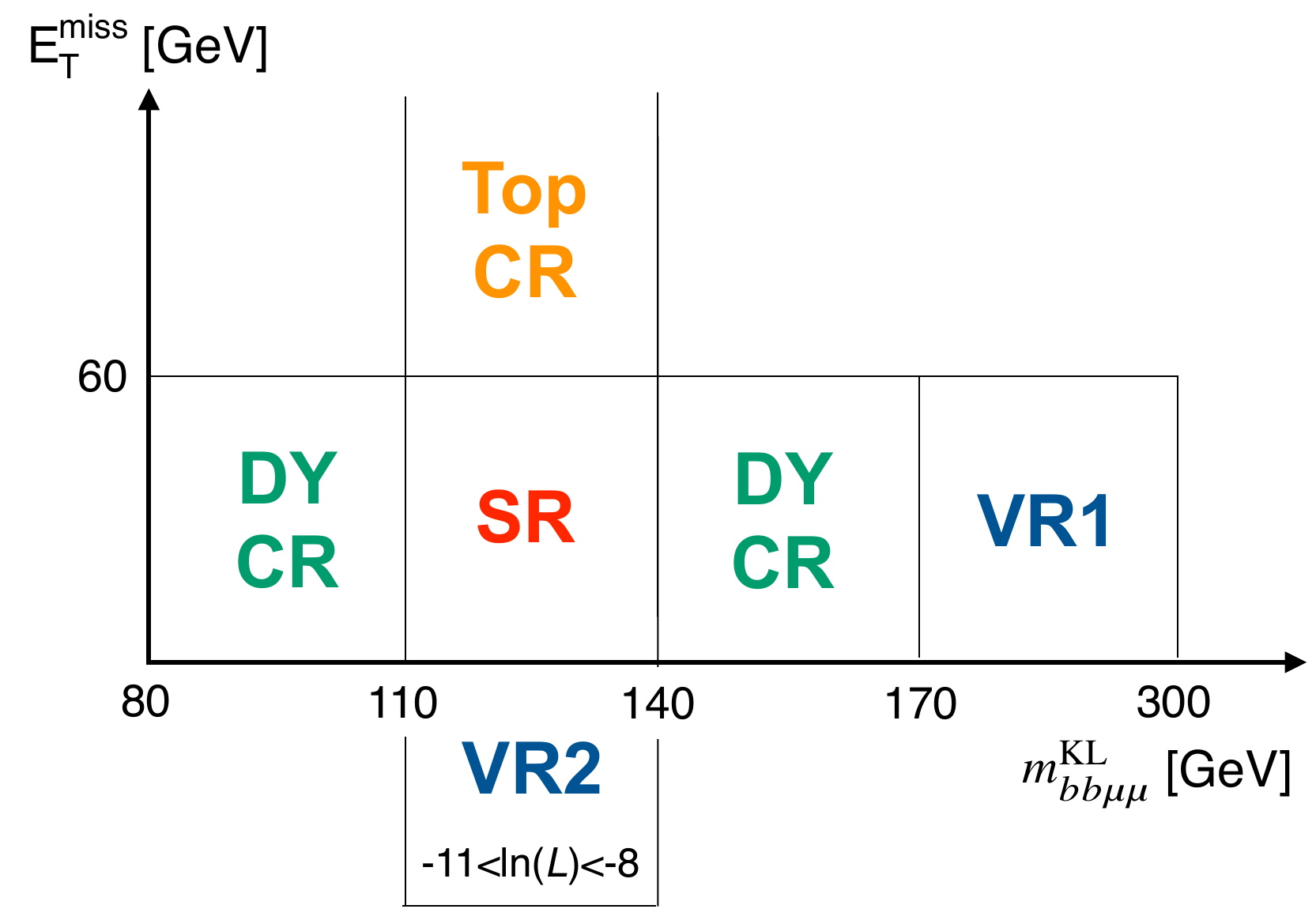}
\caption{Illustration of the signal, control and validation regions used in the analysis. The \MET requirement in the VR2 region is the same as in the SR. The corresponding DY-template regions are defined in the same way except that the two-$b$-tag requirement is changed to a zero-$b$-tag requirement.}
\label{fig:regions}
\end{figure}

The shapes of the \ttbar~kinematic variables are modelled using simulated events, while the distributions for the Drell--Yan process are taken from data templates as described below. The \ttbar~simulated sample and the DY templates are normalised in profile likelihood fits to the data. In one fit variant, the two background normalisations are simultaneously determined from the event yields in the TCR and DYCR assuming no presence of signal. 
In a second variant, the two background normalisations and the signal strength are determined using the event yields measured in the TCR, DYCR, and a given signal window.  
Two validation regions are defined to compare the number of observed events with the number of SM events predicted by the fit. One validation region (VR1) is defined in the high tail of the $bb\mu\mu$ invariant mass distribution,  $170< m_{bb\mu\mu}^{\textrm{KL}} < 300$~\GeV, while for the second validation region (VR2) only the requirement on the ln($L^\textrm{max}$) is changed relative to the SR, $-11<\text{ln}(L)<-8$. All the analysis regions are illustrated in Figure~\ref{fig:regions}.

The DY templates for each of the kinematic variables considered in a
particular region of the analysis (SR, CR or VR) are taken from the
data in a corresponding template region (DYTR). For each analysis region
the associated DYTR is defined by changing the two-$b$-tag requirement
(present in every SR, CR and VR) to a zero-$b$-tag requirement,
while keeping all other selection requirements the same. All the DYTR  are $>90$\%
pure in DY events. The small contribution from non-DY backgrounds,
namely \ttbar, dibosons, $W$+jets, single-top and $\ttbar V$, is
subtracted from the data in a DYTR using the simulated samples, and the
remaining data events are assigned to the DY template. To construct
$b$-jet-based variables, such as $m_{bb}$ and $m_{bb\mu\mu}$, in a DYTR the two leading non-tagged jets are taken and used in the computation instead of the $b$-jets.

It is verified in both the simulation and the data that the shapes of all the muon-based variables (most importantly $m_{\mu\mu}$) are consistent between the sample with no $b$-tagged jets and the sample with two $b$-tagged jets. To account for differences in jet kinematics between the DYTR dominated by light-flavour jets and the corresponding analysis region dominated by heavy-flavour jets, an event-reweighting based on the leading jet \pt\ is applied to the events in the DYTR. The event weights are derived in the data after the preselection as the ratio of the leading $b$-tagged jet \pt\ in the two-$b$-tag sample to the leading jet \pt\ in the sample with zero $b$-tags. An improvement in the modelling of jet-based kinematic variables after the reweighting is verified both in simulation and in data in the DYCR, while the shape of the $m_{\mu\mu}$ distribution remains unchanged.

Minor backgrounds include diboson production, $W$ boson production
in association with $b$-jets (with one non-prompt muon satisfying the
isolation criteria) and production of a single top quark or \ttbar\ pair
in association with a vector boson. The contribution of the minor
backgrounds in the signal region is at the percent level. They are
estimated using simulation normalised to the best available
theory prediction.

\section{Systematic uncertainties}

Dominant sources of experimental systematic uncertainty are the calibration and resolution of jet energies and muon momenta, the measurement of the $b$-tagging efficiency and the measurement of the scale and resolution of the soft term of the missing transverse momentum. Each of these uncertainties affects the \ttbar~yields by up to 14\% in any of the $m_{\mu\mu}$ bins of the signal region. 
Other experimental uncertainties have a sub-percent effect on the expected yields. These include the uncertainties in the measurement of muon identification and isolation efficiencies and the uncertainties associated with the integrated luminosity and the simulation of pile-up interactions. The uncertainty in the combined 2015+2016 integrated luminosity is 2.1\%. It is derived, following a methodology similar to that detailed in Ref.~\cite{DAPR-2013-01}, from a calibration of the luminosity scale using $x-y$ beam-separation scans performed in August 2015 and May 2016.

Four sources of theoretical uncertainty in the modelling of the \ttbar~process are considered in the analysis. As the \ttbar~simulation is normalised to the data in TCR, all of these uncertainties are applied to the acceptance ratio between TCR and SR.
Hadronisation and parton-showering model uncertainties are estimated using a sample generated with \textsc{Powheg} and showered by \textsc{Herwig++} v2.7.1 and comparing it with the nominal \textsc{Powheg} sample showered with \textsc{Pythia} v6.428.
The uncertainty due to the choice of the event generator is estimated by comparing the expected yields obtained using a \ttbar~ sample generated with \textsc{aMC@NLO} and one that is generated with \textsc{Powheg}. Both samples are showered with \textsc{Herwig++} v2.7.1.
The event generator and hadronisation/parton-showering uncertainties are found to have the largest effect among all the uncertainties affecting the total \ttbar\ expectation in the signal region: $\sim$18\% and $\sim$16\%, respectively. 
Systematic uncertainties in the modelling of initial- and final-state radiation (ISR and FSR) are assessed with \textsc{Powheg} samples showered with two alternative settings of \textsc{Pythia v6.428}. The first of these uses the PERUGIA2012radHi tune and has the renormalisation and factorisation scales set to twice the nominal value, resulting in more radiation in the final state. In addition, it has $h_{\textrm{damp}}$ set to $2 \times m_t$. The second sample, using the PERUGIA2012radLo tune, has $h_{\textrm{damp}} = m_t$ and the renormalisation and factorisation scales are set to half of their nominal values, resulting in less radiation in the event. This uncertainty has about a 5\% effect on the final \ttbar\ yields. Finally, the uncertainties due to the choice of PDF are evaluated by taking the maximum difference in the 
acceptance ratio between TCR and SR obtained with the nominal CT10 set and the alternative PDF4LHC15 set~\cite{Butterworth:2015oua}. The PDF uncertainty has up to a 2\% effect on the final \ttbar\ yields.

The uncertainties in the theoretical cross-sections (described earlier in this Letter) are assigned to the minor backgrounds whose yields are taken directly from the simulation: dibosons~(10\%), single top~(5\%) and $\ttbar V$~(13\%).
A 100\% uncertainty is applied to the $W$+jets process to account for the limited precision of the simulation when modelling the non-prompt muons satisfying the isolation criteria. Due to the minor contribution of the $W$+jets background to the analysis, this uncertainty has negligible effect. As these backgrounds have very small contributions to the SR, no theoretical uncertainties affecting the acceptance have been applied.

The systematic uncertainties applied to the data-driven DY template include the uncertainties in the shape of the template due to the 
background subtraction and different jet-flavour composition between the DYTR and SR. The uncertainty in the background subtraction is estimated from a comparison of the nominal template after the non-DY backgrounds are subtracted and the template where no subtraction is performed. The effect of this systematic uncertainty on the DY yields in the signal region is up to 4\%. 
The uncertainty in the template shape due to the jet-flavour composition is assessed by comparing the nominal template extracted from the DYTR with zero $b$-tagged jets to the template extracted from the corresponding region, but with exactly one $b$-tagged jet. The average per-bin difference between the two templates in the $m_{\mu\mu}$ distribution is taken as an overall uncertainty in the shape, amounting to 14\%.

The systematic uncertainties affecting the acceptance of the $H \rightarrow aa$ signal that correspond to the QCD scale uncertainties, the process of parton showering and hadronization and the choice of PDF set are evaluated.
The renormalisation and factorisation scales are independently varied up and down from their nominal value by a factor of two and the largest resulting change is taken as the overall uncertainty due to the QCD scale. The parton-shower uncertainties are derived by independently shifting up and down the \textsc{Pythia} internal parameters
that control the amount of ISR and FSR. Uncertainties due to the PDF are evaluated by taking the maximum difference between the yields obtained with the nominal PDF set and the alternative PDF4LHC15 and NNPDF3.0 PDF sets. 
The uncertainties due to the missing higher-order QCD corrections are applied to the ggF and VBF Higgs boson production cross-sections, amounting to 3.9\% and 2.1\%, respectively~\cite{deFlorian:2016spz,LHCHiggsCrossSectionWorkingGroup:2013}. The uncertainties due to the choice of PDF and $\alphas$ are also applied to the Higgs boson cross-section, amounting to 3.2\% for ggF and 0.4\% for VBF production~\cite{deFlorian:2016spz,LHCHiggsCrossSectionWorkingGroup:2013}.

Additionally, the ggF signal sample is compared with the alternative sample generated using the NNLOPS 
approach~\cite{Hamilton:2013fea}.
The Higgs boson rapidity distribution in the original \textsc{Powheg} signal sample is found to be consistent with the one predicted by the NNLOPS calculations, while the Higgs boson transverse momentum~($\pt(H)$) distribution is found to be harder than the one obtained using the NNLOPS approach.  A reweighting is derived as a function of $\pt(H)$ by fitting the ratio of the two  generated \pt\ distributions with a continuous function. 
The ggF signal sample is then reweighted with this function to obtain the nominal signal prediction.
A 2.5\% difference in the SR event yields observed between the weighted and unweighted sample is applied as a systematic uncertainty in the modelling of $\pt(H)$.

The signal contribution of the Higgs boson produced in the association with a vector boson ($VH$) is taken into account by increasing the total cross-section of the ggF and VBF processes by an estimated 3.5\% $VH$ contribution. A 100\% uncertainty is applied to this procedure to account for kinematic differences between the estimated $VH$ contribution and the generated ggF and VBF processes. The contribution from other Higgs boson production processes is minor and therefore not included.

Table~\ref{tab:syst} shows a summary of the dominant post-fit systematic uncertainties in the total background and signal yields across multiple $m_{\mu\mu}$ bins of the signal region. All of the uncertainties shown in Table~\ref{tab:syst}, except the normalisation and cross-section uncertainties, affect the shapes of the signal and background distributions and therefore the extrapolation of the predicted yields from the CRs to the SR.
 
\begin{table}[!htbp]
  \caption{Summary of the dominant post-fit systematic uncertainties on the background and signal yields. The uncertainties are expressed as a percentage of the total background (middle column) and signal (rightmost column) yields per $m_{\mu\mu}$ bin of the signal region. 
Shown are the uncertainties that exceed 2\% in at least one $m_{\mu\mu}$ bin. 
 \label{tab:syst}}
\begin{center}
  \begin{tabular}{l|c|c}
  Source & Total background [\%] & Signal [\%]  \\ \hline

  \hline
   DY: normalisation & 9.3--15\phantom{.} &   -- \\
   DY: flavour composition & 6.9--11\phantom{.}  &   --  \\
   DY: background subtraction & 0.4--2.4  &   --  \\
  \hline
   $\ttbar$: hard-scatter generation  & 3.6--8.6 & --  \\
   $\ttbar$: hadronisation/parton-shower &   3.2--7.7 & --  \\
   $\ttbar$: normalisation & 2.1--5.0 & --  \\
   $\ttbar$: ISR/FSR &   1.0--2.4 &  -- \\
  \hline
   MC statistics &  2.4--4.9 &  2.3--4.6 \\
   $b$-tagging &  0.6--1.5 &  17--19  \\
   Jet-energy resolution&  0.3--2.9 &  5.2--8.4 \\
   Jet-energy scale&  0.3--2.9 &  3.9--6.5  \\
   Muon-\pt resolution&  0.1--2.2 &  0.3--1.2\\ 
   Luminosity & $<$ 0.01 & 2.1 \\ 
  \hline
    Signal: QCD scale & -- & 6  \\
    Signal: ISR/FSR & -- &  4 \\
    Signal: ggF cross-section & &   \\
    \hspace{1.2cm}- missing higher-order QCD  & -- & 3.6--3.8  \\ 
    \hspace{1.2cm}- PDF \&  $\alphas$  & -- & 2.8--3.0  \\
    Signal: $VH$ contribution& -- & 3.5  \\
    Signal: $\pt(H)$ reweighting & -- & 2.3--2.5  \\
  \hline

  \end{tabular}
\end{center}
\end{table}

\section{Results}

The expected SM background in each of the analysis regions is
determined by a profile likelihood fit to the data. The numbers of
observed and predicted events in each of the bins included in the
likelihood are described by Poisson probability density functions. The systematic
uncertainties are implemented as nuisance parameters constrained by
Gaussian distributions with widths corresponding to the sizes of the
uncertainties. 

\begin{figure}[!htbp]
\captionsetup[subfloat]{farskip=-50pt,captionskip=1pt}
\centering
  \subfloat[a][]{ 
  \includegraphics[width=0.51\hsize]{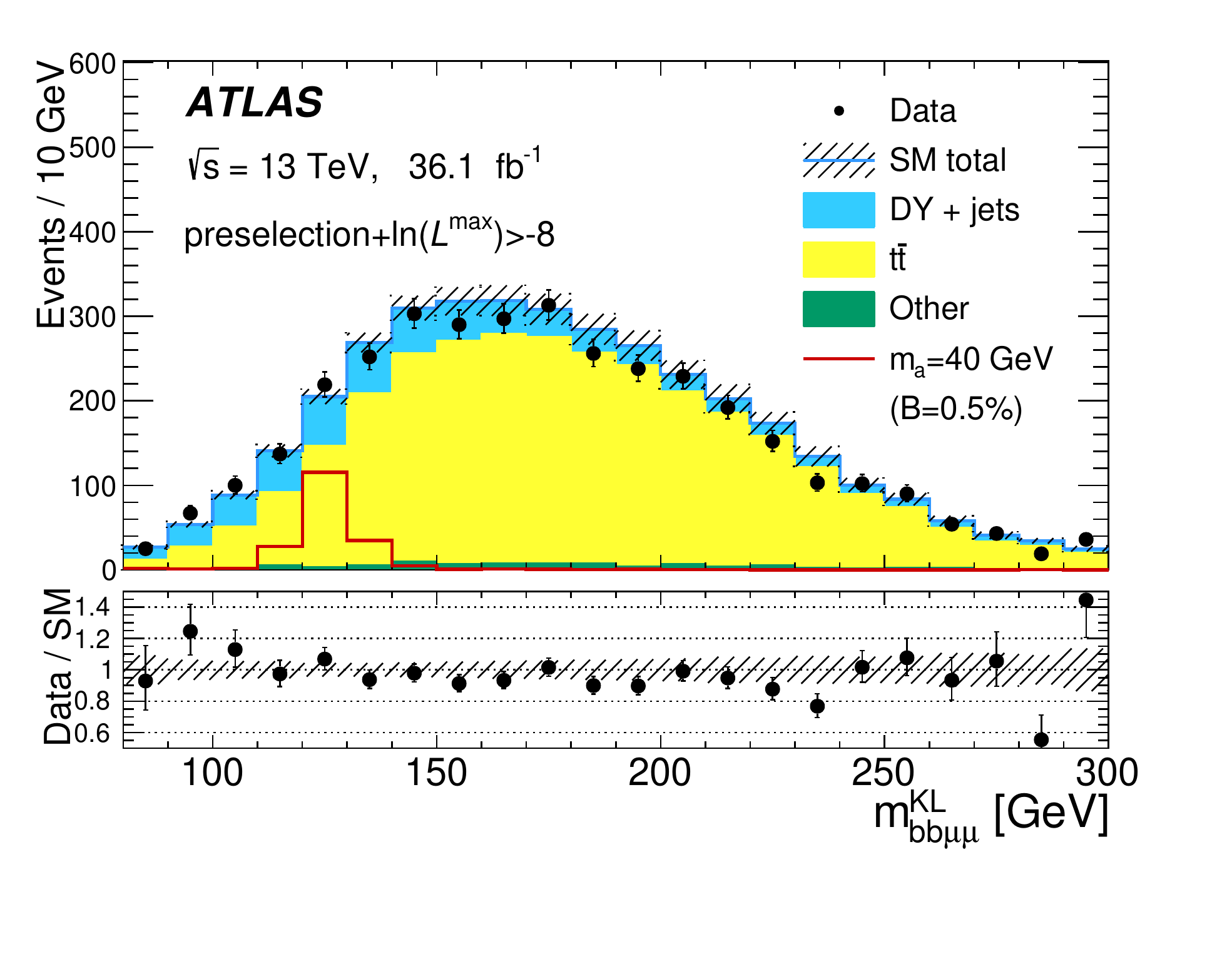} 
}
  \subfloat[b][]{
 \includegraphics[width=0.51\hsize]{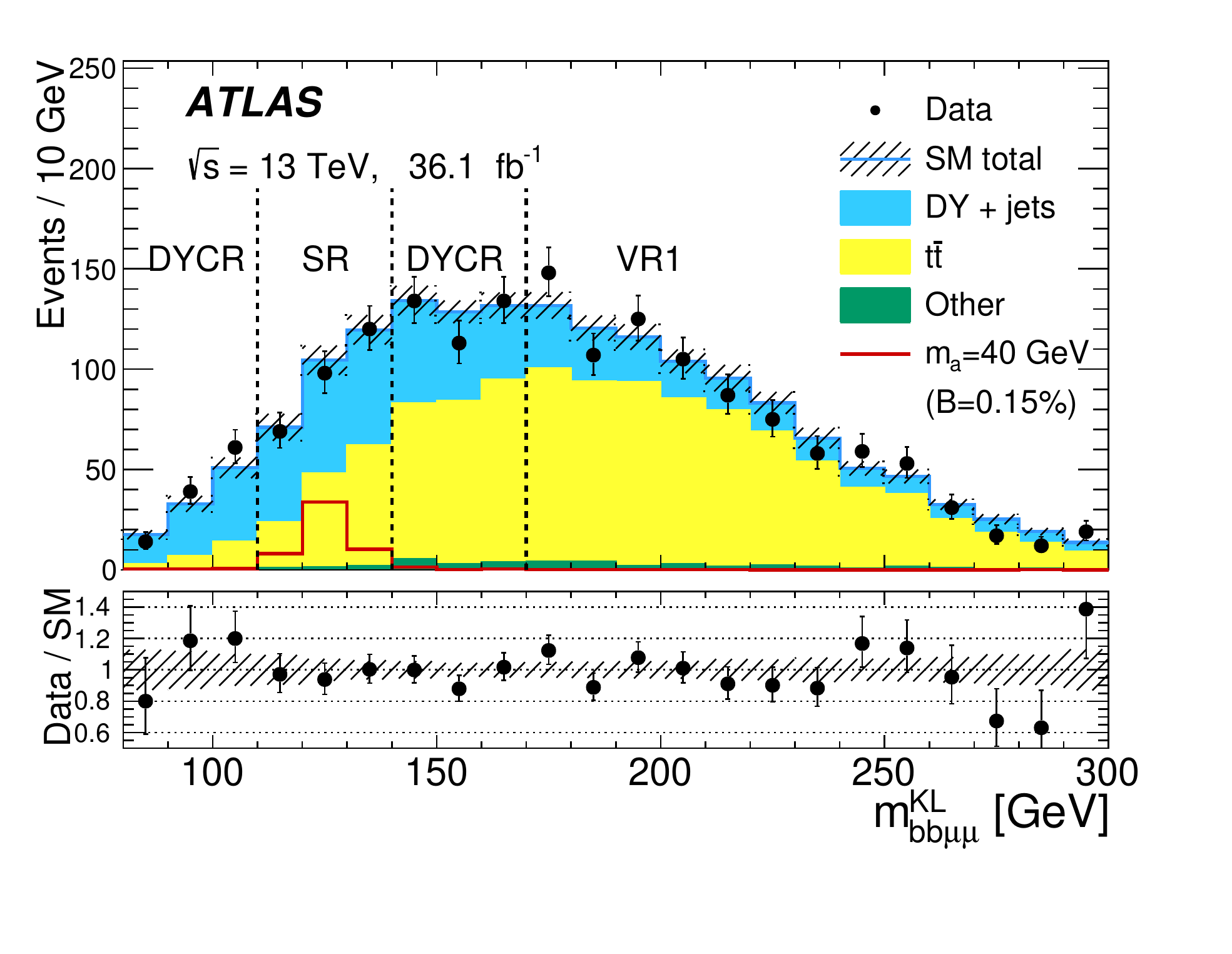}
}
\caption{The predicted and observed $m_{bb\mu\mu}^\textrm{KL}$
  distributions (a) after the preselection and the KL-fit $\ln(L^\textrm{max}) > -8$
  constraint and (b) across  DYCR, SR and VR1  (shown separated by
  vertical dashed lines). Both are shown after the background-only
  fit and differ only in the \MET< 60 GeV criterion being applied in (b).
The signal distribution for $m_a=40$~\GeV\ is also shown 
assuming the SM Higgs boson cross-section (including ggF, VBF and $VH$
production) and (a) $\mathcal{B}(H \rightarrow aa \rightarrow bb\mu\mu)=0.5\%$
and (b) $\mathcal{B}(H \rightarrow aa \rightarrow bb\mu\mu)=0.15\%$.
The hashed bands show the total statistical
and systematic uncertainties of the backgrounds.}
\label{fig:CR}
\end{figure}

\begin{figure}[!htbp]
\centering
  \includegraphics[width=0.7\hsize]{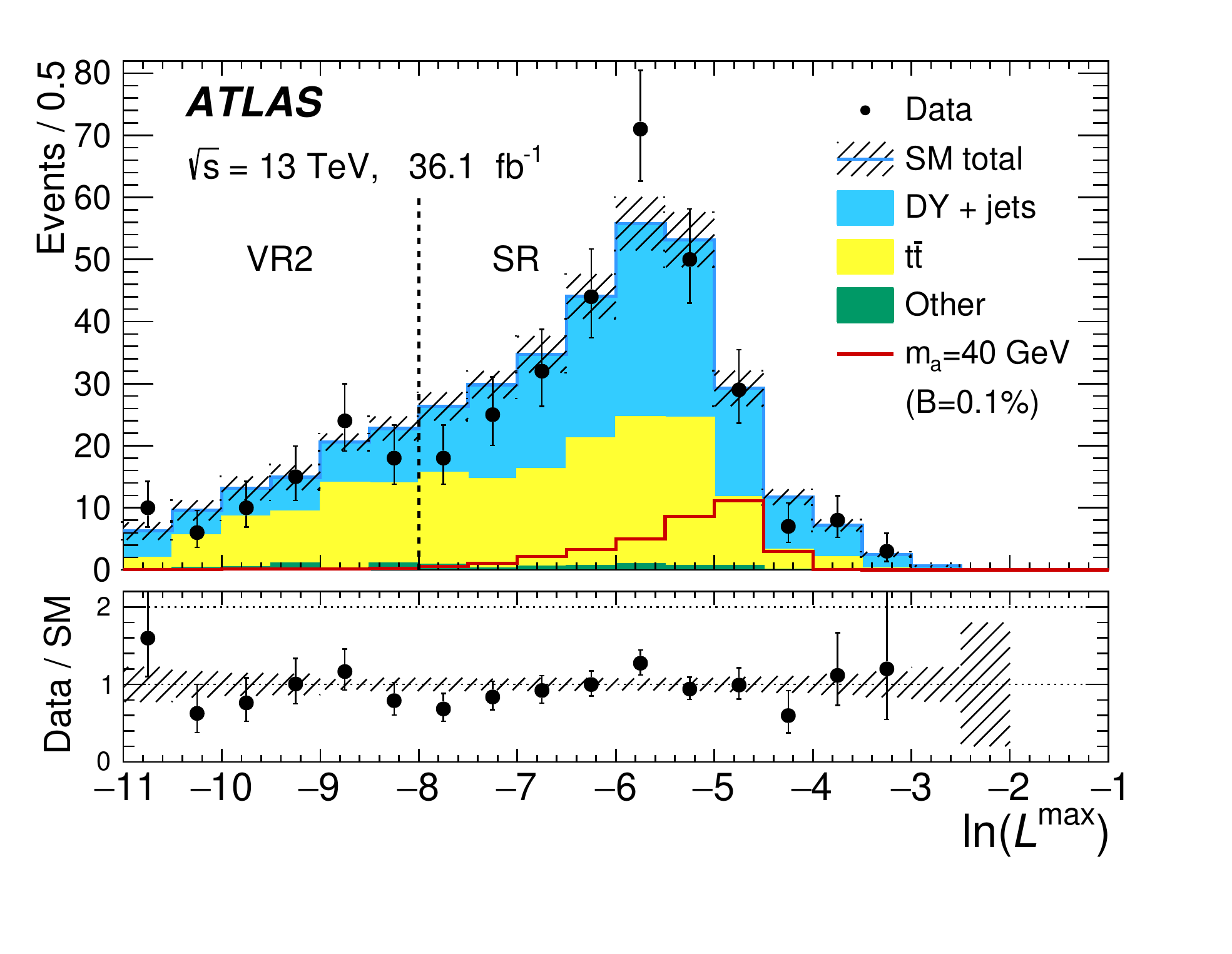}
\setlength{\abovecaptionskip}{-10pt}
\caption{The predicted and observed KL-fit $\ln(L^\textrm{max})$ distribution
  across VR2 and SR (shown separated by a vertical dashed line) after the background-only  fit. 
The signal distribution for $m_a=40$~\GeV\ is also shown 
assuming the SM Higgs boson cross-section (including ggF, VBF and $VH$
production) and $\mathcal{B}(H \rightarrow aa \rightarrow
bb\mu\mu)=0.1\%$.
The hashed bands show the total statistical
and systematic uncertainties of the backgrounds.}
\label{fig:VRlog}
\end{figure}

\begin{figure}[!htbp]
\centering
  \includegraphics[width=0.7\hsize]{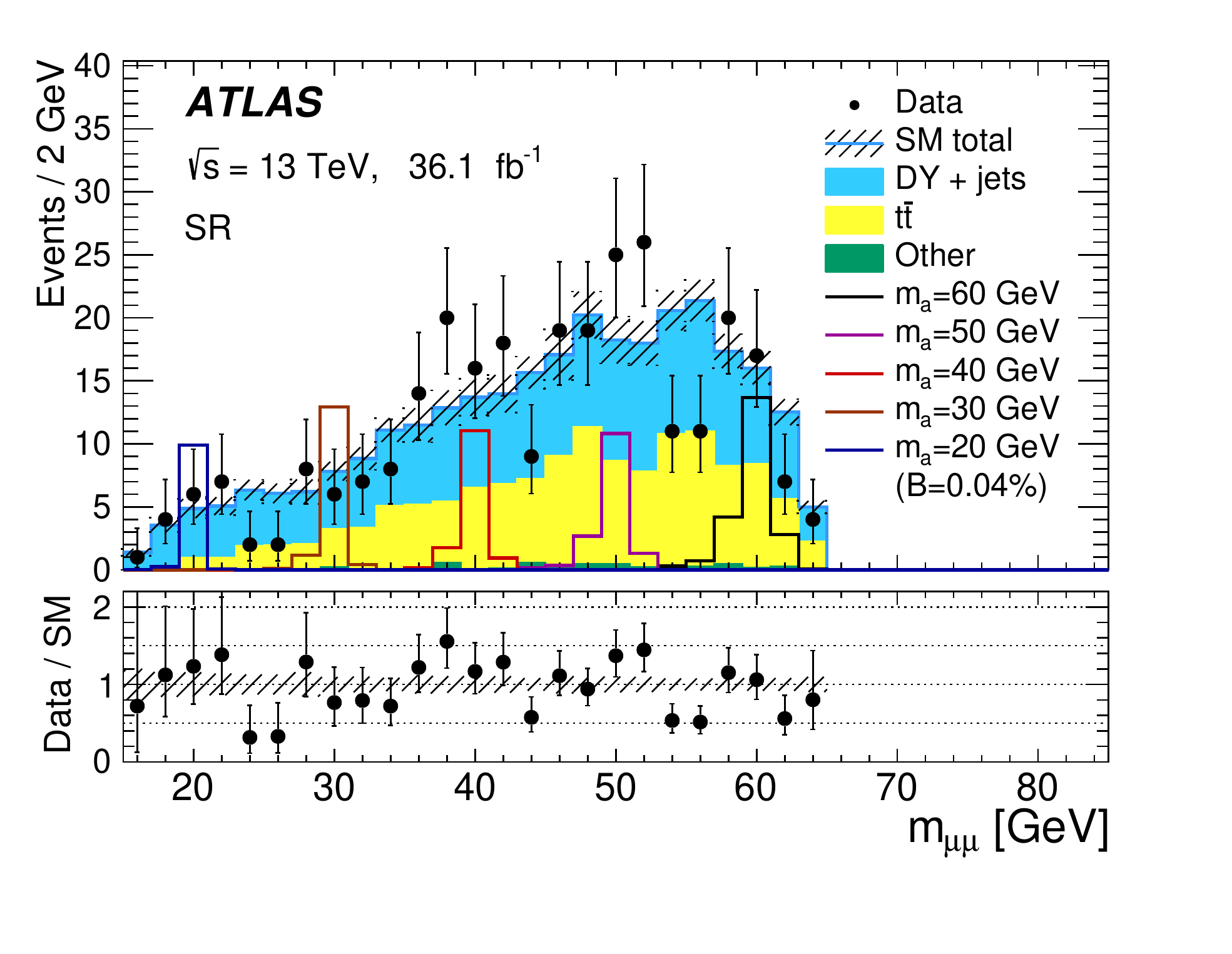}
\setlength{\abovecaptionskip}{-10pt}
\caption{The predicted and observed $m_{\mu\mu}$ 
distributions in the SR after the background-only  fit. 
The signal distributions are also shown 
assuming the SM Higgs boson cross-section (including ggF, VBF and $VH$
production) and $\mathcal{B}(H \rightarrow aa \rightarrow bb\mu\mu)=0.04\%$.
The hashed bands show the total statistical
and systematic uncertainties of the backgrounds.}
\label{fig:SR}
\end{figure}

The background-only version of the fit is performed to
verify that the post-fit background yields agree with the data
in the VRs and SR. In this version of the fit, only the data in TCR
and DYCR are used to constrain the \ttbar~and DY backgrounds and
determine their normalisation factors.  
Both TCR and DYCR are considered as one bin each.
The free fit
parameters are the overall normalisation factors
for the \ttbar~and Drell--Yan  backgrounds.  
The derived \ttbar\ (DY) normalisation factors are then applied to the number of \ttbar\ (DY)
events as predicted by the simulation (template) in any of the 
VR or SR bins. The post-fit distributions are shown in
Figures~\ref{fig:CR}--\ref{fig:SR}.
Both the normalisation and the shapes of the predicted background distributions describe the data
well in all of the analysis control and validation regions, as well as
in the SR.
The post-fit yields in five  $m_{\mu\mu}$ bins of the SR, for which the
signal sample was simulated, are shown in Table~\ref{tab:yieldsSR}. 

\begin{table}
\caption{ Total and individual background yields in five
  representative $m_{\mu\mu}$ bins of the signal region. The yields
  are the post-fit values as determined by the background-only fit. 
The uncertainties shown
  include all systematic uncertainties and the statistical MC uncertainty. $W$+jets contribution in the SR  is found to be
  negligible and is therefore not shown in the table.}
\label{tab:yieldsSR}
\begin{center}
\setlength{\tabcolsep}{0.0pc}
{\small
\begin{tabular*}{\textwidth}{@{\extracolsep{\fill}}lccccc}
\noalign{\smallskip}\hline\noalign{\smallskip}  
$m_{\mu\mu}$ bin [\GeV] &  [19--21]   & [29--31]   & [39--41]     & [48--52]    & [58--62]               \\[-0.05cm]
\noalign{\smallskip}\hline\noalign{\smallskip}
Observed events          & $6$              & $6$              & $16$              & $48$              & $29$                    \\
\noalign{\smallskip}\hline\noalign{\smallskip}
Total background         & $4.84 \pm 0.97$          & $7.8 \pm 1.2$          & $13.7 \pm 2.2\phantom{1}$          & $37.9 \pm 5.1\phantom{0}$          & $30.8 \pm 4.2\phantom{0}$              \\\noalign{\smallskip}\hline\noalign{\smallskip}
\noalign{\smallskip}\hline\noalign{\smallskip}
         \ttbar         & $0.96 \pm 0.29$          & $3.08 \pm 0.74$          & $6.6 \pm 1.5$          & $18.1 \pm 4.3\phantom{0}$          & $14.8 \pm 3.3\phantom{0}$              \\
\noalign{\smallskip}
        DY      & $3.88 \pm 0.92$          & $4.5 \pm 1.1$          & $7.1 \pm 1.7$          & $19.0 \pm 4.5\phantom{0}$          & $15.5 \pm 3.6\phantom{0}$              \\
\noalign{\smallskip}
        Dibosons     & $<$ 0.01          & $<$ 0.01          & $0.02_{-0.02}^{+0.04}\phantom{0}$          &  $0.26 \pm 0.16$          & $\phantom{0}0.3 \pm 0.1\phantom{0}$              \\
\noalign{\smallskip}
        Single top          & $<$ 0.01          & $0.2 \pm 0.2$          & $<$ 0.01          & $0.65_{-0.65}^{+0.97}\phantom{0}$          & $0.09_{-0.09}^{+0.19}\phantom{0}$              \\
\noalign{\smallskip}
        $\ttbar V$          & $<$ 0.01          &     $<$ 0.01    & $<$ 0.01          & $0.01_{-0.01}^{+0.02}\phantom{0}$          & $0.05 \pm 0.03$              \\

 \noalign{\smallskip}\hline\noalign{\smallskip}
\end{tabular*}
}
\end{center}
\end{table}

Since no significant deviation from the predicted background is
observed in the signal region,
upper limits on signal yields at 95\% confidence level (CL) 
are set as a function of $m_{\mu\mu}$ using the $\textrm{CL}_{\textrm{s}}$ prescription~\cite{Junk:1999kv,CLs_2002}. 
A series of profile likelihood fits is applied to the data in order to test 36 hypotheses for the $m_{a}$ value in steps half the size of the mass-bin width optimised in each $m_{\mu\mu}$ region. 
In each fit the likelihood function is based on the observed and predicted yields in a SR $m_{\mu\mu}$ bin corresponding to the $m_{a}$ hypothesis under test and on the expected and measured yields in the TCR and DYCR.
The profile likelihood is maximised to extract
the best-fit values for the signal strength and the \ttbar~and DY
normalisation factors.

\begin{figure}[h]
\centering
  \subfloat[a][]{
 \includegraphics[width=0.49\hsize]{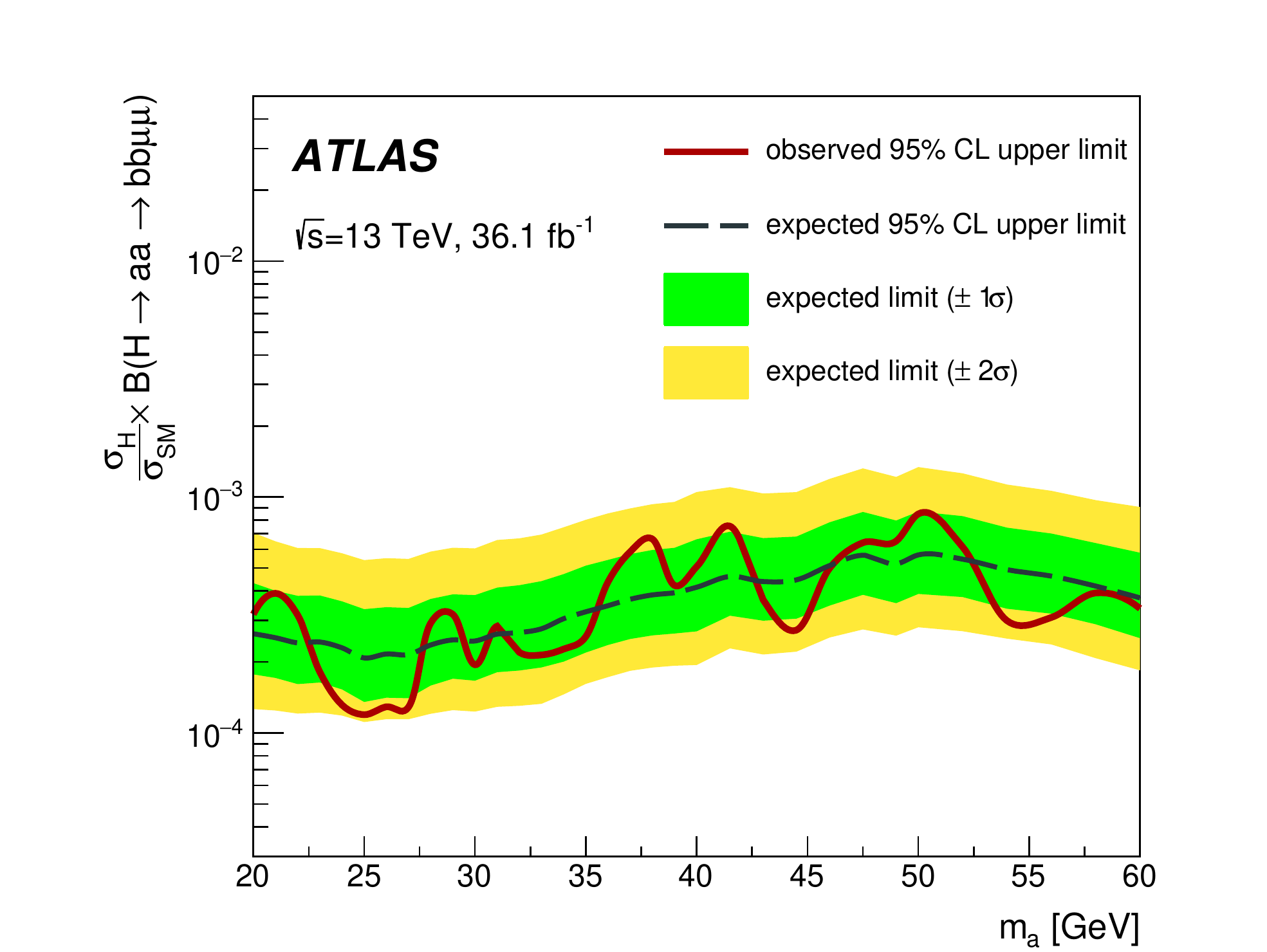}
}
  \subfloat[b][]{
 \includegraphics[width=0.49\hsize]{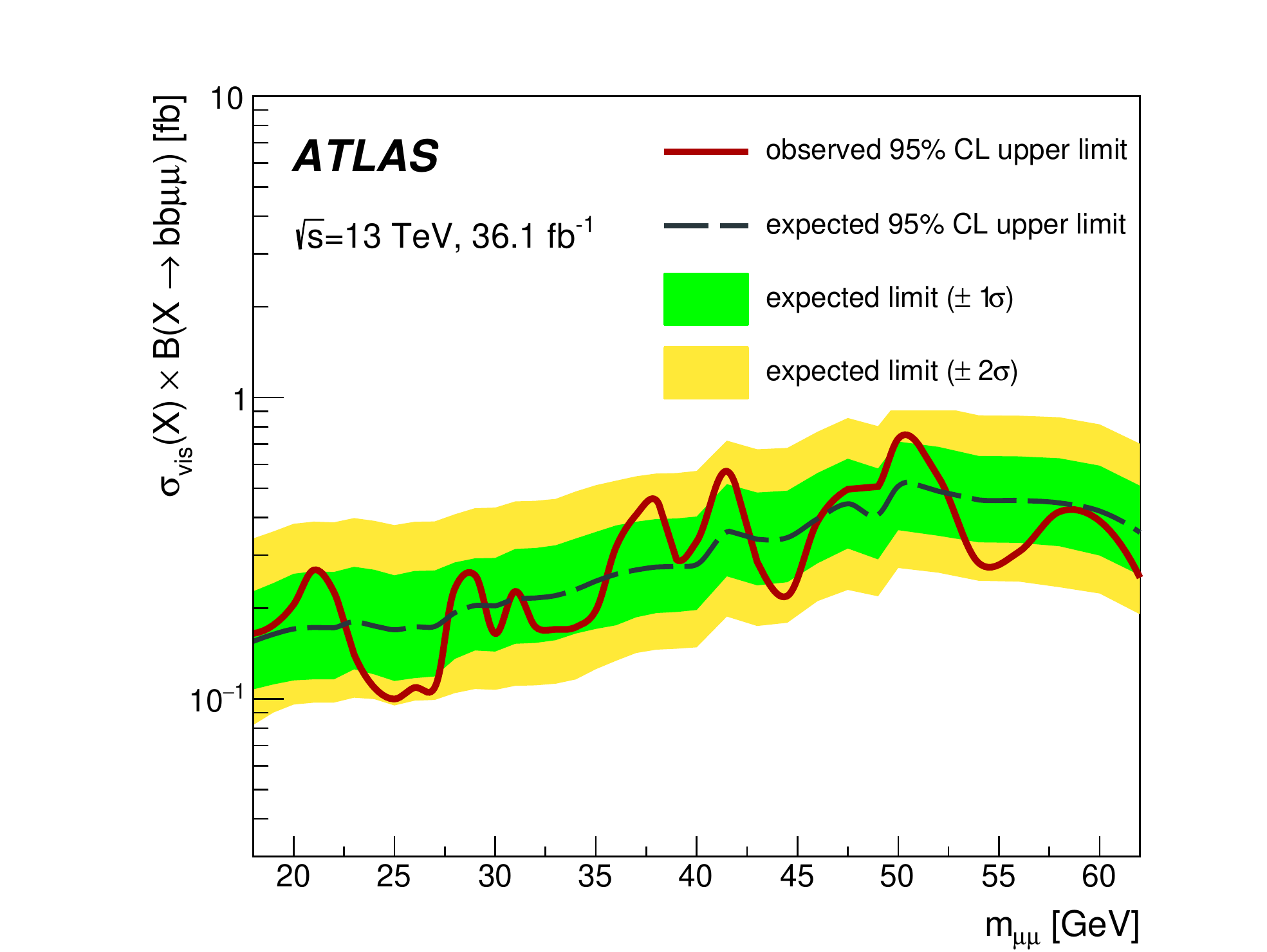}
}
\caption{ The (a) observed and expected  upper limits at the 95\%
  confidence level on $\mathcal{B}(H \rightarrow aa
  \rightarrow bb\mu\mu)$ given the SM Higgs boson production
  cross-section in the ggF, VBF and $VH$ modes and (b) model-independent upper limits on the
  visible cross-section for new physics times branching
ratio to the $bb\mu\mu$ final state $\sigma_{\textrm{vis}}(X) \times \mathcal{B}(X \rightarrow bb\mu\mu)$.}
\label{fig:limitPaper}
\end{figure}

Model-dependent limits are set on
$(\sigma_H / \sigma_{\textrm{SM}}) \times \mathcal{B}(H\rightarrow aa \rightarrow
bb\mu\mu)$ assuming the signal acceptance $\times$ efficiency as given by the
simulation. The signal acceptance $\times$
efficiency varies between 1.3\% and 2.5\% for ggF production and
between 0.94 \% and 3.2\% for VBF Higgs boson production. To obtain the
signal yield for masses for which no events were simulated, the acceptance $\times$
efficiency is interpolated with spline functions between the five simulated points.
All signal-related uncertainties are taken into account in the likelihood, with an additional
3\% interpolation uncertainty applied to the intermediate
masses. The limits are set in the $20 \leq m_a \leq 60$~\GeV\ range for which the signal
samples were simulated and range between $2 \times 10^{-4}$ and $10^{-3}$ (see Figure~\ref{fig:limitPaper}(a)).

A model-independent fit that does not include any prediction
for the signal yields in SRs and CRs is also performed. The upper limit on the number of
BSM events for each mass bin of the SR is translated to a 95\% CL upper
bound on the visible cross-section for new physics times branching
ratio into $bb\mu\mu$ final state (including the KL fit constraint
on $m_{bb} \sim m_{\mu\mu}$ and the four-object invariant mass constraint $m_{bb\mu\mu}^{\textrm{KL}} \sim m_H$), $\sigma_{\textrm{vis}}(X) \times 
\mathcal{B}(X \rightarrow bb\mu\mu)$. The visible cross-section
is defined as the product of the
production cross-section and acceptance $\times$
efficiency ($\sigma_{\textrm{vis}}(X) = \sigma_{\textrm{prod}}(X) \times
\epsilon_{\textrm{X}} $) of a potential signal after all the analysis selection criteria
have been applied. 
The limits range from 0.1~fb to 0.73~fb, depending on the dimuon mass, and are
shown in  Figure~\ref{fig:limitPaper}(b). 
The most significant excess of data over the SM prediction is found at 
$m_{\mu\mu}=38$~\GeV, with a local significance of 1.6 standard deviations.

 \FloatBarrier

\section{Conclusions}  
\label{sec:conclusions}

In summary, a search for exotic decays of the Higgs boson into two spin-zero particles
in  the $bb\mu\mu$ final state is presented. The analysis uses 36.1~fb$^{-1}$ of 
$pp$ collision data collected by ATLAS during the 2015
and 2016 runs of the LHC at $\sqrt{s}$ = 13~\TeV.  
The search for a narrow dimuon resonance is performed
over the range 18~\GeV\ $\leq m_{\mu\mu}\leq 62$~\GeV\
using mass bins that are 2, 3 or 4~\GeV\ wide depending on $m_{\mu\mu}$.
No significant excess of the data above
the SM prediction is observed. Upper limits are set on  
$(\sigma_H / \sigma_{\textrm{SM}}) \times \mathcal{B}(H\rightarrow aa\rightarrow bb\mu\mu)$ and range between $1.2 \times 10^{-4}$ and $8.4\times 10^{-4}$, depending on $m_a$. 
In Type-III 2HDM+S scenario with tan $\beta$~=~2 these limits translate into upper limits on $(\sigma_H / \sigma_{\textrm{SM}}) \times \mathcal{B}(H\rightarrow aa)$ ranging between 7\% and 47\%. The same analysis, implementing all selection criteria including $m_{bb} \sim m_{\mu\mu}$ and  $m_{bb\mu\mu}^{\textrm{KL}} \sim m_H$ constraints, is used to set the model-independent limits on the 
visible cross-section for new physics times branching ratio to the $bb\mu\mu$ final state ($\sigma_{\textrm{vis}}(X) \times 
\mathcal{B}(X \rightarrow bb\mu\mu)$),  ranging from 0.1~fb to 0.73~fb, depending on the dimuon mass.

\clearpage

\section*{Acknowledgements}

% Acknowledgements for papers with collision data
% Version 05-Oct-2018

% Standard acknowledgements start here
%----------------------------------------------
We thank CERN for the very successful operation of the LHC, as well as the
support staff from our institutions without whom ATLAS could not be
operated efficiently.

We acknowledge the support of ANPCyT, Argentina; YerPhI, Armenia; ARC, Australia; BMWFW and FWF, Austria; ANAS, Azerbaijan; SSTC, Belarus; CNPq and FAPESP, Brazil; NSERC, NRC and CFI, Canada; CERN; CONICYT, Chile; CAS, MOST and NSFC, China; COLCIENCIAS, Colombia; MSMT CR, MPO CR and VSC CR, Czech Republic; DNRF and DNSRC, Denmark; IN2P3-CNRS, CEA-DRF/IRFU, France; SRNSFG, Georgia; BMBF, HGF, and MPG, Germany; GSRT, Greece; RGC, Hong Kong SAR, China; ISF and Benoziyo Center, Israel; INFN, Italy; MEXT and JSPS, Japan; CNRST, Morocco; NWO, Netherlands; RCN, Norway; MNiSW and NCN, Poland; FCT, Portugal; MNE/IFA, Romania; MES of Russia and NRC KI, Russian Federation; JINR; MESTD, Serbia; MSSR, Slovakia; ARRS and MIZ\v{S}, Slovenia; DST/NRF, South Africa; MINECO, Spain; SRC and Wallenberg Foundation, Sweden; SERI, SNSF and Cantons of Bern and Geneva, Switzerland; MOST, Taiwan; TAEK, Turkey; STFC, United Kingdom; DOE and NSF, United States of America. In addition, individual groups and members have received support from BCKDF, the Canada Council, CANARIE, CRC, Compute Canada, FQRNT, and the Ontario Innovation Trust, Canada; EPLANET, ERC, ERDF, FP7, Horizon 2020 and Marie Sk{\l}odowska-Curie Actions, European Union; Investissements d'Avenir Labex and Idex, ANR, R{\'e}gion Auvergne and Fondation Partager le Savoir, France; DFG and AvH Foundation, Germany; Herakleitos, Thales and Aristeia programmes co-financed by EU-ESF and the Greek NSRF; BSF, GIF and Minerva, Israel; BRF, Norway; CERCA Programme Generalitat de Catalunya, Generalitat Valenciana, Spain; the Royal Society and Leverhulme Trust, United Kingdom.

The crucial computing support from all WLCG partners is acknowledged gratefully, in particular from CERN, the ATLAS Tier-1 facilities at TRIUMF (Canada), NDGF (Denmark, Norway, Sweden), CC-IN2P3 (France), KIT/GridKA (Germany), INFN-CNAF (Italy), NL-T1 (Netherlands), PIC (Spain), ASGC (Taiwan), RAL (UK) and BNL (USA), the Tier-2 facilities worldwide and large non-WLCG resource providers. Major contributors of computing resources are listed in Ref.~\cite{ATL-GEN-PUB-2016-002}.
%----------------------------------------------

\clearpage

\printbibliography

\clearpage

% ATLAS Collaboration author list
% Reference date of HIGG-2016-03 is 2018-04-18
% Author list last updated on date 11-OCT-18
% Data extracted on 11-Oct-2018 for paper reference HIGG-2016-03
% at 9:56pm
 
\begin{flushleft}
{\Large The ATLAS Collaboration}

\bigskip

M.~Aaboud$^\textrm{\scriptsize 34d}$,    
G.~Aad$^\textrm{\scriptsize 99}$,    
B.~Abbott$^\textrm{\scriptsize 125}$,    
O.~Abdinov$^\textrm{\scriptsize 13,*}$,    
B.~Abeloos$^\textrm{\scriptsize 129}$,    
D.K.~Abhayasinghe$^\textrm{\scriptsize 91}$,    
S.H.~Abidi$^\textrm{\scriptsize 164}$,    
O.S.~AbouZeid$^\textrm{\scriptsize 39}$,    
N.L.~Abraham$^\textrm{\scriptsize 153}$,    
H.~Abramowicz$^\textrm{\scriptsize 158}$
% ...
\bigskip
\\

% ...
$^{13}$Institute of Physics, Azerbaijan Academy of Sciences, Baku; Azerbaijan.\\
% ...
$^{34}$$^{(a)}$Facult\'e des Sciences Ain Chock, R\'eseau Universitaire de Physique des Hautes Energies - Universit\'e Hassan II, Casablanca;$^{(b)}$Centre National de l'Energie des Sciences Techniques Nucleaires (CNESTEN), Rabat;$^{(c)}$Facult\'e des Sciences Semlalia, Universit\'e Cadi Ayyad, LPHEA-Marrakech;$^{(d)}$Facult\'e des Sciences, Universit\'e Mohamed Premier and LPTPM, Oujda;$^{(e)}$Facult\'e des sciences, Universit\'e Mohammed V, Rabat; Morocco.\\
% ...
$^{39}$Niels Bohr Institute, University of Copenhagen, Copenhagen; Denmark.\\
% ...
$^{91}$Department of Physics, Royal Holloway University of London, Egham; United Kingdom.\\
% ...
$^{99}$CPPM, Aix-Marseille Universit\'e, CNRS/IN2P3, Marseille; France.\\
% ...
$^{125}$Homer L. Dodge Department of Physics and Astronomy, University of Oklahoma, Norman OK; United States of America.\\
% ...
$^{129}$LAL, Universit\'e Paris-Sud, CNRS/IN2P3, Universit\'e Paris-Saclay, Orsay; France.\\
% ...
$^{153}$Department of Physics and Astronomy, University of Sussex, Brighton; United Kingdom.\\
% ...
$^{158}$Raymond and Beverly Sackler School of Physics and Astronomy, Tel Aviv University, Tel Aviv; Israel.\\
% ...
$^{164}$Department of Physics, University of Toronto, Toronto ON; Canada.\\
% ...
$^{a}$ Also at Borough of Manhattan Community College, City University of New York, NY; United States of America.\\
$^{b}$ Also at California State University, East Bay; United States of America.\\
$^{c}$ Also at Centre for High Performance Computing, CSIR Campus, Rosebank, Cape Town; South Africa.\\
$^{d}$ Also at CERN, Geneva; Switzerland.\\
$^{e}$ Also at CPPM, Aix-Marseille Universit\'e, CNRS/IN2P3, Marseille; France.\\
% ...
$^{*}$ Deceased

\end{flushleft}

% Created with Glance <Atlas.Glance@cern.ch>

\end{document}